

\input phyzzx

\let\refmark=\NPrefmark 
\def\define#1#2\par{\def#1{\Ref#1{#2}\edef#1{\noexpand\refmark{#1}}}}
\def\con#1#2\noc{\let\?=\Ref\let\<=\refmark\let\Ref=\REFS
         \let\refmark=\undefined#1\let\Ref=\REFSCON#2
         \let\Ref=\?\let\refmark=\<\refsend}

\define\RWITOL
D. Olive and E. Witten, Phys. Lett. {\bf 78B} (1978) 97.

\define\RNARAIN
K. Narain, Phys. Lett. {\bf B169} (1986) 41.

\define\RNSW
K. Narain, H. Sarmadi and E. Witten, Nucl. Phys. {\bf B279} (1987) 369.

\define\RORTIN
T. Ortin, preprint SU-ITP-92-24
(hep-th/9208078).

\define\RKAL
R. Kallosh, A. Linde, T. Ortin, A. Peet and A. Van Proeyen, Phys.
Rev.{\bf D46} (1992) 5278.

\define\RSCHWARZ
J. Schwarz, preprint CALT-68-1815 (hep-th/9209125).

\define\ROLIVE
C. Montonen and D. Olive, Phys. Lett. {\bf B72} (1977) 117;

\define\ROSBORN
H. Osborn, Phys. Lett. {\bf B83} (1979) 321.

\define\RIBANEZ
A. Font, L. Ibanez, D. Lust and F. Quevedo, Phys. Lett. {\bf B249} (1990)
35;
S.J. Rey, Phys. Rev.{\bf D43} (1991) 526.

\define\RHARLIU
J. Harvey and J. Liu, Phys. Lett. {\bf B268} (1991) 40.

\define\RSTROM
A. Strominger, Nucl. Phys. {\bf B343} (1990) 167; C. Callan, J. Harvey and
A. Strominger, Nucl. Phys. {\bf B359} (1991) 611; {\bf B367} (1991) 60;
preprint EFI-91-66.

\define\RPES
I. Pesando and A. Tollsten, Phys. Lett. {\bf B274} (1992) 374.

\define\RHETEROTIC
G. Gross, J. Harvey, E. Martinec and R. Rohm, Phys. Rev. Lett. {\bf 54}
(1985) 502.

\define\RDUFF
M. Duff, Class. Quantum Grav. {\bf 5} (1988) 189;
M. Duff and J. Lu, Nucl. Phys. {\bf B354} (1991) 129, 141; {\bf B357}
(1991) 354;
Phys. Rev. Lett. {\bf 66} (1991) 1402;
Class. Quantum Grav. {\bf 9} (1991) 1;
M. Duff, R. Khuri and J. Lu, Nucl. Phys. {\bf B377} (1992) 281.
J. Dixon, M. Duff and J. Plefka, preprint CTP-TAMU-60/92
(hep-th/9208055).

\define\RGAZU
M. Gaillard and B. Zumino, Nucl. Phys. {\bf B193} (1981) 221.

\define\RODD
S. Ferrara, J. Scherk and B. Zumino, Nucl. Phys. {\bf B121} (1977) 393;
E. Cremmer, J. Scherk and S. Ferrara, Phys. Lett. {\bf B68} (1977) 234;
{\bf B74} (1978) 61;
E. Cremmer and J. Scherk, Nucl. Phys. {\bf B127} (1977) 259;
E. Cremmer and B. Julia, Nucl. Phys.{\bf B159} (1979) 141;
M. De Roo, Nucl. Phys. {\bf B255} (1985) 515; Phys. Lett. {\bf B156}
(1985) 331;
E. Bergshoef, I.G. Koh and E. Sezgin, Phys. Lett. {\bf B155} (1985) 71;
M. De Roo and P. Wagemans, Nucl. Phys. {\bf B262} (1985) 646;
L. Castellani, A. Ceresole, S. Ferrara, R. D'Auria, P. Fre and E. Maina,
Nucl. Phys. {\bf B268} (1986) 317; Phys. Lett. {\bf B161} (1985) 91;
S. Cecotti, S. Ferrara and L. Girardello, Nucl. Phys. {\bf B308} (1988)
436;
M. Duff, Nucl. Phys. {\bf B335} (1990) 610.

\define\RMAHSCH
J. Maharana and J. Schwarz, preprint CALT-68-1790 (hep-th/9207016).

\define\RSTW
A. Shapere, S. Trivedi and F. Wilczek, Mod. Phys. Lett. {\bf A6}
(1991) 2677.

\define\RBANKS
T. Banks, M. Dine, H. Dijkstra and W. Fischler, Phys. Lett.{\bf B212}
(1988) 45.

\define\RKHURI
R. Khuri, preprints CTP/TAMU-33/92 (hep-th/9205051), CTP/TAMU-35/92
(hep-th/9205081).

\define\RGAUNT
J. Gauntlett, J. Harvey and J. Liu, preprint EFI-92-67 (hep-th/9211056).

\define\RDUALITY
A. Sen, preprint TIFR-TH-92-41 (hep-th/9207053).

\define\RDYON
A. Sen, preprint TIFR-TH-92-46 (hep-th/9209016).

\overfullrule=0pt

\def\Z{Z}
\def\a{\vec\alpha}
\def\b{\vec\beta}
\def\mo{M^{(0)}}
\def\vqe{\vec q_{el}}
\def\vqm{\vec q_{mag}}
\def\lo{\lambda_1}
\def\lt{\lambda_2}
\def\loo{\lambda_1^{(0)}}
\def\lto{\lambda_2^{(0)}}
\def\vF{\vec F}
\def\tF{\tilde F}
\def\p{\partial}
\def\hG{\hat G}
\def\hB{\hat B}
\def\hA{\hat A}
\def\hqe{\hat{\vec q}_{el}}
\def\ha{\hat{\a}}
\def\haR{\ha_{0R}}
\def\haL{\ha_{0L}}
\def\ta{\tilde{\a}}

{}~\vbox{\hbox{NSF-ITP-93-29}
\hbox{TIFR-TH-93-07}\hbox{hep-th/9303057}\hbox{March,
1993}}\break

\title{MAGNETIC MONOPOLES, BOGOMOL'NYI BOUND AND SL(2,Z)
INVARIANCE IN STRING
THEORY}

\author{Ashoke Sen\foot{e-mail addresses: SEN@TIFRVAX.BITNET,
SEN@SBITP.UCSB.EDU}}

\address{Institute for Theoretical Physics, University of California,
Santa Barbara, CA 93106, U.S.A.}

\andaddress{Tata Institute of Fundamental Research, Homi Bhabha Road,
Bombay 400005, India\foot{Permanent address.}}

\abstract

We show that in heterotic string
theory compactified on a six dimensional torus, the lower bound
(Bogomol'nyi bound) on
the dyon mass is invariant under the SL(2,\Z) transformation that
interchanges
strong and weak coupling limits of the theory.
Elementary string excitations are also shown to satisfy this lower
bound.
Finally, we identify specific monopole solutions that are related
via the strong-weak coupling duality transformation to
some of the elementary particles saturating the Bogomol'nyi bound,
and these monopoles are shown to have the same mass
and degeneracy of states as the corresponding
elementary particles.

\endpage

\noindent{\bf Introduction}

Following earlier ideas\con\ROLIVE\ROSBORN\RODD\RGAZU\RSTROM\RDUFF
\RPES\RIBANEZ\RHARLIU\RSTW\noc\
we have proposed recently\RDUALITY\ that heterotic
string theory compactified on a six dimensional torus may have an SL(2,\Z)
symmetry that exchanges electric and magnetic fields, and also the strong
and weak coupling limits of the string theory.
Existence of this symmetry demands that the theory must necessarily
contain magnetically charged particles.
Allowed values of electric and magnetic charges in this theory that are
consistent with Dirac quantization condition were found, and the set of
these allowed values was shown to be invariant under SL(2,\Z)
transformation\RDYON.
This, however, does not establish that states whose quantum numbers are
related by SL(2,\Z) transformation have identical masses, $-$ a
necessary condition for SL(2,\Z) invariance of the theory.
This is the problem that we try to address in this paper.

Elementary string excitations carry only electric charge, and their
masses are well known in the weak coupling limit of the theory.
SL(2,\Z) transform of these states carry both electric and magnetic
charges in general, and must arise as soliton solutions in this theory.
Thus in order to establish the SL(2,\Z) invariance of the mass spectrum,
we must compare the elementary particle masses at weak coupling to the
soliton masses at strong coupling.
In a generic theory, calculating soliton masses at strong coupling would
have been an impossible task; however, since the theory under
consideration has $N=4$ supersymmetry, one can derive some
results about the soliton masses in this theory that are not
expected to receive any quantum corrections\RWITOL.
In particular, for a soliton carrying a given amount of electric and
magnetic charges, one can derive a lower bound (known as the
Bogomol'nyi bound) for the mass of the soliton.
The bound is saturated for supersymmetric solitons, and the masses of
such solitons are expected not to receive any quantum corrections.
Thus one can compare these exact mass formulae as well as the lower
bound on the soliton masses with the masses of the elementary string
excitations and ask if they agree with the postulate of SL(2,\Z)
invariance of the theory.
Although this would not prove that SL(2,\Z) is a symmetry of the theory,
this would provide a stringent test of this symmetry.

In this paper we show first that the Bogomol'nyi bound is invariant under
SL(2,\Z) transformation, and second, that the masses of the elementary
string excitations also satisfy the Bogomol'nyi bound, with a subset of
them saturating the bound.
This implies that the elementary string excitations saturating the
Bogomol'nyi bound, and the supersymmetric solitons whose quantum numbers
are related to those of these elementary particles by SL(2,\Z)
transformation, have the same mass.
We also identify the specific soliton solutions that are
related by an SL(2,\Z) transformation to some of the elementary string
excitations saturating the Bogomol'nyi bound.

Some other aspects of SL(2,\Z) invariance have been discussed in
ref.\RSCHWARZ.

\noindent {\bf Review}

The low energy effective action describing ten dimensional heterotic
string  theory is given  by
$$\eqalign{
S =& {1\over 32\pi} \int d^{10} x \sqrt{-\det G^{(10)}_S} e^{-\Phi^{(10)}}
\Big(R^{(10)}_S + G_S^{(10)MN} \p_M\Phi^{(10)}
\p_N\Phi^{(10)}\cr
& -{1\over 12}
G_S^{(10)MM'} G_S^{(10)NN'} G_S^{(10)TT'} H^{(10)}_{MNT} H^{(10)}_{M'
N'T'}
 -{1\over 8} G_S^{(10)MM'} G_S^{(10)NN'} F^{(10)I}_{MN} F^{(10)I}_{M'
N'}\Big)\cr
}
\eqn\eone
$$
where
$$
F^{(10)I}_{MN}=\p_M A^{(10)I}_N - \p_N A^{(10)I}_M
\eqn\ethree
$$
and
$$
H^{(10)}_{MNT}=\Big( \p_M B^{(10)}_{NT} -{1\over 4} A^{(10)I}_M
F^{(10)I}_{NT} +{\rm ~ cyclic ~ permutations ~ of ~} M,N,T \Big)
\eqn\etwo
$$
Here $\Phi^{(10)}$ is the dilaton field, $G^{(10)}_{SMN}$ denote ten
dimensional $\sigma$-model metric,
$B^{(10)}_{MN}$ denote the rank two antisymmetric tensor field, and
$A^{(10)I}_M$ denote 16 $U(1)$ gauge fields.
The superscript $^{(10)}$ indicates that we are dealing with ten
dimensional fields, the indices $M, N, T$ are ten dimensional Lorentz
indices and run from 0 to 9, and the indices $I$ denote 16 dimensional
gauge indices and run from 1 to 16.
Note that we have included only the abelian gauge fields in the effective
action.
For a generic toroidal compactification to four dimensions, all the
non-abelian symmetry is spontaneously broken, and only the $U(1)$
gauge fields remain massless\RNARAIN\RNSW.

We now compactify the theory on a 6 dimensional torus.
Let us denote by $m,n$ ($1\le m,n\le 6$) the six internal directions,
and by $\mu, \nu$ ($\mu, \nu=0, 7, 8, 9$) the four uncompactified
directions.
In terms of the ten dimensional fields, we define the four dimensional
fields as follows:\foot{In writing down these relations, we have made
a change of normalization from the one used in ref.\RDUALITY\ to the
one used in ref.\RDYON.}
$$\eqalign{
& \hG_{mn} = G^{(10)}_{Smn}, ~~~ \hB_{mn}= B^{(10)}_{mn}, ~~~
\hA^I_m = A^{(10)I}_m, ~~~ \Phi = \Phi^{(10)} -{1\over 2} \ln\det\hG \cr
& A^m_\mu ={1\over 2} \hG^{mn} G^{(10)}_{Sn\mu}, ~~~
A_\mu^{I+12}= -{1\over 2\sqrt 2} A^{(10)I}_\mu +{1\over\sqrt 2}\hA^I_m
A^m_\mu,\cr
& A_\mu^{m+6}={1\over 2} B^{(10)}_{m\mu}-\hB_{mn} A^n_\mu
+{1\over 2\sqrt 2}\hA^I_m A^{I+12}_\mu\cr
&G_{S\mu\nu}=G^{(10)}_{S\mu\nu}-G^{(10)}_{Sm\mu} G^{(10)}_{Sn\nu}
\hat G^{mn}, ~~~
B_{\mu\nu}= B^{(10)}_{\mu\nu}-4 B^{(10)}_{mn}A^m_\mu A^n_\nu
-2 (A^m_\mu A^{m+6}_\nu -\mu\leftrightarrow \nu)\cr
& \qquad \qquad 1\le m,n \le 6, ~~~ 1\le I\le 16\cr
}
\eqn\efour
$$
where $\hG^{mn}$ denotes the inverse matrix of $\hG_{mn}$.
The field strengths associated with the
four dimensional gauge fields and the anti-symmetric
tensor field are defined as
$$
F^\alpha_{\mu\nu}= \p_\mu A^\alpha_\nu - \p_\nu A^\alpha_\mu, \qquad
1\le \alpha\le 28
\eqn\efive
$$
and
$$
H_{\mu\nu\rho}=\Big( \p_\mu B_{\nu\rho} + 2 A^\alpha_\mu
L_{\alpha\beta} F^\beta_{\nu\rho} +{\rm ~ cyclic ~ permutations ~ of~}
\mu, \nu, \rho\Big)
\eqn\esix
$$
where $L$ denotes the $28\times 28$ matrix,
$$
L=\pmatrix{ 0 & I_6 & 0\cr I_6 & 0 & 0\cr 0 & 0 & -I_{16}\cr}
\eqn\ethirteena
$$
The Einstein metric in 4 dimensions is obtained from the metric
$G_{S\mu\nu}$ through the rescaling,
$$
G_{\mu\nu} = e^{-\Phi}G_{S\mu\nu}
\eqn\eseven
$$
{}From now on, we shall choose the convention that all four dimensional
indices will be raised and lowered with the Einstein metric.
With this convention, we define the dual field strength
$$
\tF^{\alpha\mu\nu}={1\over 2} (\sqrt{-\det G})^{-1}
\epsilon^{\mu\nu\rho\sigma} F^\alpha_{\rho\sigma}
\eqn\eeight
$$

The equations of motion of the anti-symmetric tensor field allow us to
define a scalar field $\Psi$ through the equation:
$$
H^{\mu\nu\rho}= -(\sqrt{-\det G})^{-1} e^{2\Phi} \epsilon^{\mu\nu\rho
\sigma}\p_\sigma\Psi
\eqn\enine
$$
We can combine the fields $\Phi$ and $\Psi$ into a single complex scalar
field $\lambda$:
$$
\lambda=\Psi + i e^{-\Phi}\equiv \lambda_1 + i\lambda_2
\eqn\eten
$$
Finally, all information about the scalar fields $\hG_{mn}$, $\hB_{mn}$
and $\hA^I_m$ may be included in a single 28 dimensional matrix $M$
satisfying,
$$
M^T=M, ~~~ M^T L M=L
\eqn\ethirteen
$$
$M$ is defined as,
$$
M=\pmatrix{P & Q & R\cr Q^T & S & U\cr R^T & U^T & V\cr}
\eqn\eeleven
$$
where,
$$\eqalign{
P^{mn}=&\hG^{mn}, ~~~ Q^m_{~n}=\hG^{mp}(\hB_{pn}+{1\over 4} \hA^I_p
\hA^I_n), ~~~ R^{mI}={1\over\sqrt 2}\hG^{mp}\hA^I_p,\cr
S_{mn}=&(\hG_{mp}-\hB_{mp}+{1\over 4}\hA^I_m\hA^I_p) \hG^{pq}
(\hG_{qn}+\hB_{qn} +{1\over 4}\hA^J_q\hA^J_n)\cr
U^{~I}_m=& {1\over\sqrt 2}(\hG_{mp} -\hB_{mp}+{1\over 4}\hA^J_m
\hA^J_p)\hG^{pq}\hA^I_q, ~~~
V^{IJ}=\delta^{IJ}+{1\over 2}\hA^I_p\hG^{pq}\hA^J_q\cr
}
\eqn\etwelve
$$
and $^T$ denotes the transpose of a matrix.
The equations of motion derived from the action \eone\ can be shown
to be equivalent to those derived from the action\RMAHSCH\RDYON
$$\eqalign{
S =& {1\over 32\pi} \int d^4 x \sqrt{-\det G}\Big( R - {1\over 2
(\lambda_2)^2} G^{\mu\nu}\p_\mu\lambda \p_\nu\bar\lambda
-\lambda_2\vec F^T_{\mu\nu} . LML . \vF^{\mu\nu}\cr
&+\lambda_1 \vF^T_{\mu\nu}.L.\tilde{\vF}^{\mu\nu}
+{1\over 8} G^{\mu\nu} Tr(\p_\mu M L \p_\nu M L)\Big)\cr
}
\eqn\efourteen
$$
where we have used vector notation to denote the 28 dimensional vector
$F^\alpha_{\mu\nu}$.
The equations of motion derived from the above action may be shown to
be invariant under the SL(2,\Z) transformation\RSTW\RDUALITY
$$\eqalign{
& \lambda\to {a\lambda +b\over c\lambda + d}, ~~~
G_{\mu\nu}\to G_{\mu\nu}, ~~~ M\to M, ~~~
\vF_{\mu\nu}\to (c\lo +d)\vF_{\mu\nu} + c\lt ML .\tilde{\vF}_{\mu
\nu}\cr
&\qquad\qquad a,b,c,d\in \Z, ~~~ ad -bc=1\cr
}
\eqn\efifteen
$$

The electric and magnetic charges of a particle are defined in terms
of the asymptotic values of the electric and magnetic fields as
follows
$$
\vF_{0r}\simeq {\vqe\over r^2}, ~~~ \vec{\tF}_{0r}\simeq {\vqm\over
r^2}
\eqn\esixteen
$$
where $r=\sqrt{(x^7)^2+(x^8)^2+(x^9)^2}$.
The allowed spectrum of $(\vqe,\vqm)$ in toroidally compactified
heterotic string theory was calculated in ref.\RDYON, and was
found to be of the form
$$
\vqm=\mo L\b, ~~~ \vqe={1\over \lto}(\a + \loo\b)
\eqn\eseventeen
$$
where $\loo$, $\lto$ and $\mo$ denote the asymptotic values of
$\lambda_1$, $\lambda_2$ and $M$ respectively, and $\a$ and $\b$ are
arbitrary vectors belonging to a lattice $P$ which is even and
self-dual with respect to the metric $L$.

\noindent{\bf Bogomol'nyi Bound and its SL(2,\Z) Invariance}

An explicit formula for the Bogomol'nyi bound on the mass of a dyon
for toroidally compactified heterotic string theory was derived in
ref.\RHARLIU.\foot{Lower bound to magnetically charged black hole mass in
supersymmetric theories was derived in ref.\RKAL.
Invariance of the Bogomol'nyi bound for the mass of dyonic black
holes under SL(2,R) transformation was shown in ref.\RORTIN.}
We shall first write down this formula, then reexpress it in terms of
the charges $\vqe$, $\vqm$ defined through eqs.\esixteen, and finally
show that it is invariant under the SL(2,\Z) transformation of
$\vqe$, $\vqm$ induced from eq.\efifteen.

Let us define,
$$
T_{(m)\mu\nu}=\p_\mu G^{(10)}_{m\nu}-\p_\nu G^{(10)}_{m\mu}
-H^{(10)}_{m\mu\nu}
\eqn\eeighteen
$$
We define $\tilde T_{(m)\mu\nu}$ through an equation analogous to
eq.\eeight\ and then define the charges $Q_m$, $P_m$ through the
asymptotic values of these fields as follows:
$$
T_{(m)0r}\simeq {Q_m\over r^2}, ~~~ \tilde T_{(m)0r}\simeq
{P_m\over r^2}
\eqn\enineteen
$$
In terms of these charges, the Bogomol'nyi bound on the dyon mass may
be expressed as\RHARLIU
$$
m^2\ge {1\over 64}\lto (\hG^{(0)mn}Q_m Q_n +\hG^{(0)mn} P_m P_n)
\equiv (m_0)^2
\eqn\etwenty
$$
provided $G_{\mu\nu}$ approaches $\eta_{\mu\nu}$ asymptotically.
Here the superscript $^{(0)}$ denotes asymptotic values of various
fields.
Using eqs.\efour, and the definition of $Q_m$, $P_m$ given in
eqs.\eeighteen, \enineteen, we get,
$$\eqalign{
Q_m=& 2q_{el}^{m+6} + 2 (\hG^{(0)}_{mn} +\hB^{(0)}_{mn}
+{1\over 4} \hA^{(0)I}_m \hA^{(0)I}_n)q^n_{el} -\sqrt 2
\hA^{(0)I}_m q^{I+12}_{el}\cr
P_m=& 2q_{mag}^{m+6} + 2 (\hG^{(0)}_{mn} +\hB^{(0)}_{mn}
+{1\over 4} \hA^{(0)I}_m \hA^{(0)I}_n)q^n_{mag} -\sqrt 2
\hA^{(0)I}_m q^{I+12}_{mag}\cr
}
\eqn\etwentyone
$$
Substituting this in eq.\etwenty\ we get the following expression for
the Bogomol'nyi bound $m_0$:
$$
(m_0)^2={\lto\over 16}\{ \vqe^T.(L\mo L +L). \vqe + \vqm^T .(L\mo L+L).
\vqm\}
\eqn\etwentytwo
$$
Before testing SL(2,\Z) invariance of $m_0$, let us note that for states
carrying $q^\alpha_{el}$, $q^\alpha_{mag}$ charge 0 for $1\le\alpha\le 12$,
we get,
$$
(m_0)^2={\lto\over 32}\{
(\hA^{(0)I}_p q_{el}^{I+12})\hG^{(0)pq} (\hA^{(0)J}_q q^{J+12}_{el})
+
(\hA^{(0)I}_p q_{mag}^{I+12})\hG^{(0)pq} (\hA^{(0)J}_q q^{J+12}_{mag})\}
\eqn\etwentythree
$$
This is precisely Osborn's formula\ROSBORN\ for the Bogomol'nyi bound
on the monopole mass for a global $N=4$ supersymmetric Yang-Mills
theory. The fields $\hA^I_m$ should be interpreted as Higgs fields
in this case.\foot{All comparisons are made in the gauge where the
asymptotic Higgs field is directed along a fixed direction in the
gauge group, except along a Dirac string singularity.}

Let us now study the SL(2,\Z) transformation law of $(m_0)^2$.
Using eqs.\efifteen, \esixteen\ we see that under SL(2,\Z)
transformation,
$$\eqalign{
\lto &\to {\lto\over |c\lambda^{(0)} +d|^2}\cr
\vqe &\to (c\loo +d)\vqe + c\lto\mo L.\vqm\cr
\vqm &\to (c\loo + d)\vqm -c\lto\mo L.\vqe\cr
}
\eqn\etwentyfour
$$
Using eqs.\ethirteen\ and \etwentytwo\ we get
$$
m_0\to m_0
\eqn\etwentyfive
$$
under the transformation \etwentyfour.
The above result implies that if we find two states whose quantum
numbers are related by SL(2,\Z) transformation, and if both of these
states saturate the Bogomol'nyi bound, then their masses are
automatically identical.

\endpage

\noindent{\bf Where do Known Monopole Solutions Fit in?}

Eq.\eseventeen\ gives the allowed spectrum of electric and magnetic charges
of a monopole.
We shall now try to analyse the asymptotic fields of various known
monopole solutions\RBANKS\RHARLIU\RKHURI\RGAUNT\ and see where they fit
in this list.
Eq.\eseventeen, however, is not the most convenient starting point for this
analysis, since the lattice $P$, to which the vectors $\a$ and $\b$
belong, itself depends on $\mo$.
In particular, if $P_0$ denotes the lattice $P$ for $\mo=I$, then from
eqs.\efourteen, \esixteen\ and \eseventeen\ we see that,
$$
P=(L\mo L)^{-1}P_0=\mo P_0
\eqn\etwentysix
$$
where we have used eq.\ethirteen\ to get the last relation in
eq.\etwentysix.
Using eq.\ethirteen\ we see that $P_0$ is also an even, self-dual,
Lorentzian lattice with metric $L$.
We can now express $\a$ and $\b$ as,
$$
\a =\mo \a_0, ~~~ \b=\mo \b_0, ~~~~~\a_0, \b_0\in P_0
\eqn\etwentyseven
$$
Eq.\eseventeen\ may now be rewritten as,
$$
\vqm=L\b_0, ~~~~ \vqe={1\over\lto}\mo (\a_0+\loo\b_0)
\eqn\etwentyeight
$$
The lattice $P_0$ to which $\a_0$ and $\b_0$ belong is now independent
of $\mo$.
We shall call $\a_0$ and $\b_0$ electric and magnetic charge vectors
respectively.

Let us now consider the BPS monopole solution in string theory
discussed in refs\RBANKS\RHARLIU\ in the gauge where asymptotically
the Higgs field is directed along a fixed direction in the gauge space,
except along a Dirac string singularity.
With the normalization convention that we have chosen, the asymptotic
values of various fields are given by,
$$\eqalign{
& B^{(10)}_{\mu\nu}\simeq 0, ~~~ G^{(10)}_{S\mu\nu}\simeq
Diag(-1, e^{2\phi_0},
e^{2\phi_0}, e^{2\phi_0}), ~~~
\Phi^{(10)}\simeq 2\phi_0, ~~~
\p_\mu G^{(10)}_{m\nu} -\p_\nu
G^{(10)}_{m\mu}\simeq 0\cr
& H^{(10)}_{m0 r}\simeq 0, ~~~ H^{(10)}_{mij}\simeq 8C\delta_{m,1}
\epsilon_{ijk}{x^k\over r^3}, ~~~ F^{(10)I}_{0r}\simeq 0, ~~~
F^{(10)I}_{ij}\simeq -4\delta^{I,1}\epsilon_{ijk}{x^k\over r^3}\cr
& B^{(10)}_{mn}\simeq 0, ~~~ A^{(10)I}_m\simeq 4C\delta_{m,1}
\delta^{I,1},
{}~~~ G^{(10)}_{Smn}\simeq Diag(e^{2\phi_0}, 1, 1, 1, 1, 1)\cr
& \qquad \qquad 7\le i, j\le 9, ~~ 1\le m, n\le 6, ~~ 1\le I\le 16\cr
}
\eqn\etwentynine
$$
Using eq.\efour\ we see that $\phi_0$ denotes the asymptotic value
of the four dimensional dilaton field $\Phi$.
Using eqs.\eseven\ and \eten\ we get,
$$
G_{\mu\nu}\simeq Diag(-e^{-\phi_0}, e^{\phi_0}, e^{\phi_0}, e^{\phi_0}),
{}~~~ \lto=e^{-\phi_0}
\eqn\ethirty
$$
We now scale the internal coordinate $x^1$ by $e^{\phi_0}$, the time
coordinate $x^0$ by $e^{-\phi_0/2}$, and the space coordinates
$x^7$, $x^8$, $x^9$ by $e^{\phi_0/2}$, so that asymptotically
$G^{(10)}_{Smn}$ approaches $\delta_{mn}$ and $G_{\mu\nu}$
approaches $\eta_{\mu\nu}$.
In this new coordinate system the various transformed fields are
given by,
$$\eqalign{
& B^{(10)}_{\mu\nu}\simeq 0, ~~~ G^{(10)}_{S\mu\nu}\simeq
Diag(-e^{\phi_0}, e^{\phi_0},
e^{\phi_0}, e^{\phi_0}), ~~~
\Phi^{(10)}\simeq 2\phi_0, ~~~ \p_\mu G^{(10)}_{m\nu} -\p_\nu
G^{(10)}_{m\mu}\simeq 0\cr
& H^{(10)}_{m0 r}\simeq 0, ~~~ H^{(10)}_{mij}\simeq 8C\delta_{m,1}
e^{-\phi_0}\epsilon_{ijk}{x^k\over r^3}, ~~~ F^{(10)I}_{0r}\simeq 0, ~~~
F^{(10)I}_{ij}\simeq -4\delta^{I,1}\epsilon_{ijk}{x^k\over r^3}\cr
& B^{(10)}_{mn}\simeq 0, ~~~ A^{(10)I}_m\simeq 4C
e^{-\phi_0}\delta_{m,1}\delta^{I,1},
{}~~~ G^{(10)}_{Smn}\simeq \delta_{mn}\cr
}
\eqn\ethirtyone
$$
Using eqs.\efour\ and \ethirtyone\ we can find the asymptotic values of
various four dimensional fields.
Here we only list those which are asymptotically non-trivial:
$$\eqalign{
F^\alpha_{0r}\simeq & 0, ~~~
\tF^{\alpha}_{0r}\simeq \delta_{\alpha,13}
{\sqrt 2\over r^2}\cr
\hA^I_m\simeq & 4C\delta^{I,1} \delta_{m,1} e^{-\phi_0}, \qquad
\qquad 1\le \alpha\le 28, ~~1\le m\le 6, ~~ 1\le I\le 16\cr
}
\eqn\ethirtytwo
$$
In particular, note that non-trivial $H^{(10)}_{1ij}$ is induced
solely by $\tilde F^{13}_{0r}$.
Comparing with eq.\esixteen\ we get,
$$
\vqe=0, ~~~ q^\alpha_{mag}=\delta^{\alpha, 13}\sqrt 2
\eqn\ethirtythree
$$
Since these solutions do not carry any electric charge,
they are valid solutions only
for $\lo=0$.\foot{Dyon solutions in this theory can also be
constructed\RHARLIU.}
Comparison with eq.\etwentyeight\ gives,
$$
\a_0=0, ~~~ \beta^\alpha_0=-\delta^{\alpha, 13} \sqrt 2
\eqn\ethirtyfour
$$
Note that $\a_0$ and $\b_0$ are even with respect to the inner
product metric $L$, as is required by the quantization condition.
Also note that here allowed values of $\a_0$ and $\b_0$ are
quantized, but the constant $C$ is arbitrary.

Next let us turn to the $H$ monopole solutions\RBANKS\RKHURI\RGAUNT\ for
which all fields become asymptotically trivial except for the
field strength associated with the antisymmetric tensor field.
The only non-trivial asymptotic field component is given by,
$$
H^{(10)}_{mij}\simeq Q\delta_{m,1} \epsilon_{ijk} {x^k\over r^3}
\eqn\ethirtyfive
$$
where $Q$ is some parameter.
Using eqs.\efour, \esixteen, and \etwentyeight\ we get,
$$
\a_0=0, ~~~ \beta_0^\alpha=-{1\over 2} Q\delta_{\alpha, 1}
\eqn\ethirtyseven
$$
Again notice that $\a_0$ and $\b_0$ are even with respect to the inner
product metric $L$ (in fact both $\a_0^T.L.\a_0$ and $\b_0^T.L.\b_0$
vanish.)
The requirement that $\b_0$ lies on the lattice $P_0$ gives
rise to the quantization condition on $Q$, as discussed in ref.\RGAUNT.

\noindent{\bf Where do Elementary String Excitations Fit in?}

We now try to see whether the elementary string excitations satisfy the
Bogomol'nyi bound, and, if they do, then which are the excitations that
saturate the bound.
Since elementary string excitations do not carry any magnetic charge,
we rewrite eq.\etwentytwo\ for particles carrying electric charge only:
$$
(m_0)^2={\lto\over 16} \vqe^T. (L\mo L+L). \vqe
\eqn\ethirtyeight
$$
We can simplify the above expression by using the observation of ref.\RNSW\
that the physics remains invariant under a simultaneous change of
the background $\mo$ and the lattice of electric charge vectors of
the form:
$$
\mo\to \Omega \mo \Omega^T, ~~~ \vqe\to \Omega\vqe
\eqn\ethirtyeighta
$$
with $\Omega$ satisfying,
$$
\Omega L \Omega^T = L
\eqn\ethirtyeightb
$$
Under this transformation
$$
P\to \Omega P, ~~~ P_0\to L\Omega L P_0
\eqn\ethirtyeightc
$$

Let us choose $\Omega=\Omega_0$ such that $\Omega_0\mo\Omega_0^T=I$,
and denote all the transformed variables by putting a hat on top of
them.
In this case,
$$\hat \mo =I
\eqn\ethirtyeighte
$$
and,
$$
\hqe={1\over \lto}\ha_0, ~~~ \ha_0
\equiv L\Omega_0 L\a_0 \in \hat P_0\equiv L\Omega_0 L P_0
\eqn\ethirtyeightf
$$
Eq.\ethirtyeight\ may now be rewritten as,
$$
(m_0)^2={\lto\over 16} \hqe^T .(I+L) . \hqe ={1\over 8\lto}
(\haR)^2
\eqn\ethirtyeightg
$$
where,
$$
\hat{\a}_{0{R\atop L}}\equiv {1\over 2} (I\pm L)\hat {\a}_0
\eqn\ethirtyeighth
$$

We now turn to
the mass formula for elementary string excitations\RHETEROTIC.
This takes a simple form in terms of the vector $\hat{\a}_0$:
$$
m^2={1\over 8\lto} \{(\haR)^2+2N_R-1\} ={1\over 8\lto}
\{(\haL)^2+2N_L-2\}
\eqn\efortythree
$$
In the above expression $(\haR)^2$ and $(\haL)^2$ denote the
internal momentum contributions, $N_R$ and $N_L$ denote the oscillator
contributions, and $-1$ and $-2$ denote the ghost contributions in the
right and the left sectors respectively.
(In our notation the right hand sector is the world-sheet
supersymmetric sector.)
GSO projection requires $N_R$ to be at least $1/2$, since we need
a factor of $\psi^M_{-1/2}$ to create the lowest mass state in the
Neveu-Schwarz sector.\foot{Since the Ramond sector states are degenerate
with the Neveu-Schwarz sector states, we do not need to analyze
the mass formula in the Ramond sector separately.}
Eq.\efortythree\ then gives,
$$
m^2\ge {1\over 8\lto}(\haR)^2
\eqn\efortyfour
$$
which is the same bound as eq.\ethirtyeightg.
The elementary particle states saturating the Bogomol'nyi bound have
$N_R=1/2$, but, as we can see from eq.\efortythree, $N_L$ is not
fixed for these states.

\noindent{\bf Monopole Solutions Conjugate to Elementary
Particles Saturating the Bogomol'nyi Bound}

We shall now indicate how to identify
the monopole solutions which have
quantum numbers related
via the SL(2,\Z) transformation $\lambda\to -
1/\lambda$
to those of the elementary particles saturating
the Bogomol'nyi bound.
Invariance of the Bogomol'nyi bound under SL(2,\Z) transformation will
then automatically tell us that these states have the same mass.
We concentrate on the transformation $\lambda\to -1/\lambda$,
since this transformation sends $\ha_0$ to $-\hat{\b}_0$
and $\hat{\b}_0$ to $\ha_0$\RDYON, and hence,
acting on the purely electrically charged states, produces purely
magnetically charged states for $\loo=0$.

We have seen that the elementary string excitations saturating the
Bogomol'nyi bound has $N_R=1/2$, but $N_L$ is unrestricted.
We shall now analyze the three cases separately: $N_L=0$, $N_L=1$ and
$N_L\ge 2$.

\noindent Case I: $N_L=0$. Here
$$
m^2={1\over 8\lto}(\haR)^2 ={1\over 8\lto}\{(\haL)^2 -2\}
\eqn\efortysix
$$
so that,
$$
(\hat{\a})^2\equiv (\haL)^2 - (\haR)^2=2
\eqn\efortyseven
$$
In this case the only ten dimensional Lorentz index of the state
comes from the oscillator $\psi^M_{-1/2}$ in the right hand sector.
Together with the Ramond sector states, these states form massive
vector supermultiplets of $N=4$ supersymmetry algebra.
We shall now show that each of these states may be interpreted as
belonging to an $SU(2)$ gauge multiplet, that has become massive
due to the spontaneous breaking of the $SU(2)$ gauge symmetry.
To do this we again use the trick of changing the lattice $P_0$ at the
cost of changing the background $\mo$ as described in eq.\ethirtyeighta.
This time we look for a transformation matrix $\omega$ such that,
$$
\ta_0\equiv L\omega L\hat{\a}_0
\eqn\efortyeight
$$
has,
$$
\ta_{0R}\equiv {1\over 2}(I+L)\ta_0=0
\eqn\efortynine
$$
We define,
$$
\tilde \mo=\omega\omega^T
\eqn\efifty
$$
so that eq.\efortysix\ may now be rewritten as,
$$
m^2={1\over 16\lto} \ta^T_0.(\tilde \mo +L).\ta_0
\eqn\efiftyone
$$
We can interpret
$\ta_0$ as the new electric charge vector
lying on the lattice $\tilde P_0= L\omega L
\hat P_0$, and $\tilde \mo$ as the new background value of $M$.

Let us now note that if, keeping the lattice $\tilde P_0$ fixed, we
had set $\tilde \mo$ to $I$, then $m^2$ would vanish.
This, in turn, shows that these states may be interpreted as otherwise
massless states, which have acquired mass due to the background
$\tilde\mo$.
More specifically, these states may be interpreted as $SU(2)$ gauge
bosons and their superpartners, which have acquired mass due to
spontaneous breaking of SU(2) by the background $\tilde\mo$.
The charged generators of this $SU(2)$ group correspond to the
vectors $\pm\ta_0$ on the lattice $\tilde P_0$.

The monopoles related to these charged particles
by $\lambda\to -1/\lambda$ transformation are characterized by
zero electric charge vector, and magnetic charge vector $\ta_0$.
These are precisely the BPS monopoles associated with the spontaneous
breaking of this particular $SU(2)$, constructed in
refs.\RBANKS\RHARLIU.
It is also known\ROSBORN\RHARLIU\ that these monopoles belong to the
massive vector supermultiplet of the $N=4$ supersymmetry algebra.
Hence
we see that there is an exact one to one correspondence between the
elementary particle states corresponding to $N_R=1/2$, $N_L=0$, and
the monopole states whose quantum numbers are
related to these by the SL(2,\Z) transformation
$\lambda\to -1/\lambda$.
The masses of these monopoles and the elementary particle states are
also identical, since they both saturate the Bogomol'nyi bound, and
this bound has already been shown to be invariant under the SL(2,\Z)
transformation.

We should note, however, that the solutions in refs.\RBANKS\RHARLIU\ are
constructed as a power series expansion in the scale of breaking of the
SU(2) symmetry, which, in the present case, corresponds to a power
series expansion in $(\ha_{0R})^2$.
Thus, the explicit form of the solution can be written down only for
small $(\ha_{0R})^2$; but we expect that the general features of the
solution, e.g.
partially unbroken supersymmetry, mass, and the degeneracy of states, will
remain unchanged even for finite $(\ha_{0R})^2$.

\noindent Case II: $N_L=1$. In this case eq.\efortythree\ takes the
form:
$$
m^2={1\over 8\lto}(\ha_{0R})^2={1\over 8\lto}(\ha_{0L})^2
\eqn\efiftytwo
$$
so that,
$$
(\ha_0)^2\equiv (\ha_{0L})^2-(\ha_{0R})^2=0
\eqn\efiftytwo
$$
Thus the conjugate monopoles in this case will be characterized by zero
electric charge vector and magnetic charge vector $\ha_0$ with
zero norm.
Although one to one correspondence between monopoles and elementary
particle states has not been established in this case, there are
certainly known examples of such monopoles.
These are the $H$-monopole solutions (monopoles carrying purely
anti-symmetric tensor field charge) discussed in
eqs.\ethirtyfive, \ethirtyseven,
and also the Kaluza-Klein type of monopoles which are related to these
$H$-monopole solutions via the usual $R\to 1/R$
(or more general $O(6, 22; \Z)$) duality transformation\RBANKS.

\noindent Case III: $N_L\ge 2$. In this case,
$$
m^2 ={1\over 8\lto} (\ha_{0R})^2 ={1\over 8\lto} \big( (\ha_{0L})^2
+2N_L -2\big)
\eqn\efiftyfive
$$
so that,
$$
(\ha_0)^2=(\ha_{0L})^2 -(\ha_{0R})^2=2-2N_L\le -2
\eqn\efiftysix
$$
The monopoles conjugate to these have magnetic charge vector $\ha_0$
with negative norm with respect to the metric $-L$.
There are no known monopole solutions with this quantum number.
This, however, is not surprising, since, as we shall argue now,
construction of such monopole solutions will probably involve massive
string fields in a non-trivial way.
To see this, let us note that in the two previous cases, there is a
limit ($(\ha_{0R})^2\to 0$) in which the monopole mass vanishes.
Such monopoles must be constructed out of nearly massless fields.
On the other hand, in this case, there is no limit in which the monopole
is massless, since from eq.\efiftyfive\ we see that $m^2\ge (1/4\lto)$.
Hence there is no reason why one should be able to construct
such solutions purely in terms of nearly  massless fields.
Thus it appears that the only way to construct these monopole solutions
would be to look for exact conformal field theories.

\noindent {\bf Conclusion}

To summarize, in this paper we have identified the monopole
solutions related via the SL(2,\Z) transformation $\lambda\to -1/
\lambda$ to some of the elementary string excitations, and have shown
that these monopoles have the same mass and degeneracy of states as
the elementary string excitations. Furthermore, we have shown that
both, the dyon solutions, and the elementary excitations in string
theory satisfy a lower bound to their masses, and this lower bound
is invariant under the SL(2,\Z) transformation.
These results provide a further support to the conjecture that SL(2,\Z)
might be an exact symmetry of heterotic string theory compactified on
a six dimensional torus.

\ack This work was supported in part by the National Science
Foundation grant no. PHY89-04035.

\refout

\end